# Mesoporous bioactive glass/ε-polycaprolactone scaffolds promote bone regeneration in osteoporotic sheep

N. Gómez-Cerezo[a,b], L. Casarrubios[c], M. Saiz-Pardo[d], L. Ortega[d], D. de Pablo[d], I. Díaz-Güemes[e], B. Fernández-Tomé[e], S. Enciso[e], F. M. Sánchez-Margallo[e], M.T. Portolés[c*], D. Arcos[a,b*], M. Vallet-Regí[a,b,*]

[a] *Departamento de Química en Ciencias Farmacéuticas, Facultad de Farmacia, Universidad Complutense de Madrid, Instituto de Investigación Sanitaria Hospital 12 de Octubre i+12, Plaza Ramón y Cajal s/n, 28040 Madrid, Spain*

[b] *CIBER de Bioingeniería Biomateriales y Nanomedicina (CIBER-BBN), Spain*

[c] *Departamento de Bioquímica y Biología Molecular, Facultad de Ciencias Químicas. Universidad Complutense de Madrid, Instituto de Investigación Sanitaria Hospital Clínico San Carlos (IdISSC), 28040 Madrid, Spain*

[d] *Servicio de Anatomía Patológica. Hospital Clínico San Carlos, Facultad de Medicina Universidad Complutense de Madrid, Instituto de Investigación Sanitaria Hospital Clínico San Carlos (IdISSC), 28040 Madrid, Spain*

[e] *Centro de Cirugía de Mínima Invasión Jesús Usón, Cáceres, Spain*

\* Corresponding authors
*E-mail address:* portoles@quim.ucm.es, arcosd@ucm.es, vallet@ucm.es

Phone: +34 91 394 4666; Fax: +34 91 394 4159

**Abstract**

Macroporous scaffolds made of a $SiO_2$-$CaO$-$P_2O_5$ mesoporous bioactive glass (MBG) and ε-polycaprolactone (PCL) have been prepared by robocasting. These scaffolds showed an excellent *in vitro* biocompatibility in contact with osteoblast like cells (Saos 2) and osteoclasts derived from RAW 264.7 macrophages. *In vivo* studies were carried out by implantation into cavitary defects drilled in osteoporotic sheep. The scaffolds evidenced excellent bone regeneration properties, promoting new bone formation at both the peripheral and the inner parts of the scaffolds, thick trabeculae, high vascularization and high presence of osteoblasts and osteoclasts. In order to evaluate the effects of the local release of an antiosteoporotic drug, 1% (%wt) of zoledronic acid was incorporated to the scaffolds. The scaffolds loaded with zoledronic acid induced apoptosis in Saos 2 cells, impeded osteoclast differentiation in a time dependent manner and inhibited bone healing, promoting an intense inflammatory response in osteoporotic sheep.

# 1. Introduction

Osteoporosis has been defined as "a systemic disease characterized by low bone mass and micro-architectural deterioration of bone tissue, with consequent increase in bone fragility and susceptibility to fracture" [1] and it is associated with significant morbidity and mortality [2,3]. Osteoporosis also hinders the clinical success of endosseous implants and grafting materials for the treatment of bone defects [4]. The systemic metabolism and the site morphological conditions make difficult not only the primary stability, but also biological fixation, osteointegration and bone regeneration processes [5,6].

Mesoporous bioactive glasses (MBGs) are a group of bioceramics that have arisen great interest in the field of regenerative medicine of bone [7,8] and for local drug delivery purposes [9,10]. These materials combine the bone regenerative properties of bioactive glasses [11,12] with the high surface area and porosity of silica based mesoporous materials [13]. The synergy between chemical composition and textural properties of MBGs results in bone grafts with excellent bioactive behavior under both *in vitro* [14–16] and *in vivo* [17,18] conditions. In combination with biocompatible polymers, MBGs can be fabricated as scaffolds for bone tissue regeneration purposes [19-21], resulting in porous structures that exhibit, excellent biocompatibility, good differentiation markers activities in preosteoblast and mesenchymal stem cells, capability to host osteogenic agents and other therapeutics and the improvement of mechanical properties of the macroporous structures [22,23]. In this sense, the robocasting method (also referred as 3D printing) has been widely used for manufacturing MBG based scaffolds with highly regular and interconnected macroporosity [24-29]. Among biocompatible polymers, ε-polycaprolactone (ε-PCL) has shown very interesting features, especially when is manufactured as 3D scaffolds using additive manufacturing techniques [30-32]. More specifically, ε-PCL has been widely used as continuous phase for MBG-PCL composites manufactured by robocasting, as PCL exhibits plastic properties that facilitate its processing by extrusion through the narrow nozzles used in robocasting techniques [33-36]. Moreover, ε-PCL acts a binder, supplying the required rheological characteristics for the obtention of structural stability of the struts and layers during the printing. Besides, the robocasting of MBG-PCL mixtures produces scaffolds with uniform pore structures able to stimulate the cell colonization by enhancing osteoblast proliferation and differentiation [37], thus being excellent candidates for bone grafting purposes, even in osteoporotic patients. However, no *in vivo* evidence has been provided about the real potential of this combination.

The association of drugs with bone grafting materials has provided very positive outcomes for the treatment of bone defects [38-40]. In this sense, zoledronic acid is an antiresorptive bisphosphonate that inhibits the osteoclastic activity, increases mineral bone density (MBD), enhances compression strength of bone and significantly decreases the risk of fractures. Based on these clinical outcomes, several authors have proposed the incorporation of zoledronic acid in bone grafts, with the aim of improving the outcomes of several biomaterials in bone healing processes [41-48]. However, administration of zoledronic acid leads to a decrease of bone remodeling markers, including those associated to bone formation such as collagen type I and alkaline phosphatase [49]. Besides, it has been suggested, but not demonstrated, that zoledronic acid interferes with the reparative mechanisms of bone and leads to adynamic bone disease [42]. Identifying the positive or deleterious effects of zoledronic acid on bone regeneration requires *in vivo* studies in appropriated animal models. These studies should simulate the osteoporosis conditions in humans, using biomechanically relevant animal models.

The main aim of this work has been the preparation of MBG-PCL macroporous scaffolds for bone regeneration in an osteoporotic sheep model. In order to determine the beneficial or deleterious effect of zoledronic acid incorporated to these scaffolds. To achieve these aims, highly bioactive MBG-PCL macroporous scaffolds were designed and fabricated by robocasting and implanted in osteoporotic sheep, which is a clinically relevant model of human post-menopausal bone loss and comparable to human osteoporosis in terms of osteoporosis degree, deterioration of trabecular structure, bone metabolism, hormone profile and anatomy [50–53].

## 2. Materials and methods

*2.1. Synthesis of mesoporous bioactive glass MBG-58S*

Mesoporous bioactive glass MBG-58S with composition 58 $SiO_2$ - 37 CaO - 5 $P_2O_5$ (% mol) was synthesized by evaporation induced self-assembly (EISA) process [54,55], using the nonionic surfactant Pluronic F127 as structure directing agent. F127 (2 g) were dissolved in 38 g of ethanol with 1 mL of 0.5 M HCl solution. Afterwards, 3.75 mL of tetraethyl orthosilicate (TEOS), 0.44 mL of triethyl phosphate (TEP) and 2.47 g of $Ca(NO_3)_2 \cdot 4H_2O$ were added under continuous stirring in 3 h intervals. The obtained sols were cast in Petri dishes (9 cm diameter) to undergo the EISA process at 30 °C for 7 days in Petri dishes and the resulting gels were treated at 700 °C for 3 h to obtain the final MBG-58S as a calcined glass powder. The powder was sieved and the particles size fraction between 25 and 40 micrometers was collected. All reactants were purchased from Sigma-Aldrich and used without further purification.

*2.2. Preparation of mesoporous bioactive glass/ε-polycaprolactone (MBG-PCL) scaffolds*

ε-polycaprolactone (3 g) of Mw = 58,000 Da (Sigma-Aldrich) were dissolved in 40 mL of dichloromethane. Subsequently 4.5 g of MBG-58S suspended in 37.5 mL of dichloromethane were incorporated to this solution and mixed until obtaining a homogeneous suspension. In order to facilitate the extrusion of this mixture through our robotic deposition apparatus, the solvent was partially evaporated for 2 hours at room temperature. For comparison purposes, scaffolds containing zoledronic acid (MBG-PCL-zol) were prepared including 75 mg of zoledronic acid (United States Pharmacopeia, USP, Reference Standard) in the mixture, which correspond to 1 % in weight. In human adults, zoledronic acid is commonly administered intravenously as a 4-5 mg/100 mL infusion once a year. Since the average weight of each scaffold is about 415 mg, we decided to incorporate 1% in weight, i.e. 4.15 mg of zoledronic acid, similarly to the dosage used in humans.

Macroporous scaffolds were prepared by robocasting of this mixture using a robotic deposition apparatus (EnvisionTEC GmbH Prefactory VR 3-D BioplotterTM). Previously, a stl model was designed as a cylinder with 14 mm in diameter, as previous experiments showed a volume reduction of 28 % in diameter after robocasting during the drying of the scaffolds. The scaffolds were printed by depositing the paste through a cylindrical nozzle (diameter = 460 μm) at the volumetric flow rate required to maintain a constant x–y table speed of 5 mm·s$^{-1}$ with a height increase between layers along z axis of 450 μm. Each scaffold was robocasted with 25 layers, reaching an initial height of 11.5 mm. A constant temperature of 25 ºC was kept in the paste reservoir and in the deposition plate. Finally, the scaffolds were kept at 40ºC overnight to remove the rest of remaining dichloromethane. Cylindrical scaffolds with diameter of 10 mm, height of 10 mm and 415 (± 12) mg in weight, were obtained after complete precipitation and drying.

*2.3. Physico-chemical characterization*

Textural properties of the scaffolds were determined by nitrogen adsorption with a Micromeritics 3Flex equipment. Samples were previously degassed under vacuum for 24 h, at 40 °C. Differential thermal analysis was carried out using a TG/DTA Seiko SSC/5200 thermobalance between 50 ºC and 1000 °C at a heating rate of 1 ºC·min$^{-1}$. Both, nitrogen adsorption and thermogravimetric analysis were carried out in triplicate.

The scaffolds surfaces were studied by Fourier transform infrared (FTIR) spectroscopy using a Nicolet Magma IR spectrometer and by scanning electron microscopy (SEM) using a JEOL F-6335 microscope, operating at 20 kV and equipped with an energy dispersive X-ray spectrometer.

*2.4. Study of zoledronic acid release*

Since zoledronic acid is included in the raw ε-polycaprolactone-MBG mixture before printing the scaffolds, *in vitro* drug delivery assays were performed on the assumption that the whole amount of zoledronic acid is incorporated into the scaffolds, in the same manner than the other components, i. e. MBG and ε-polycaprolactone. Previously, the calibration curve for zoledronic acid in a phosphate buffer saline (PBS) at pH 7.4 and 37 °C was calculated by measuring the absorbance at 209 nm of different solutions with concentrations between 0.001 mg/mL and 0.15 mg/mL. MBG-PCL-zol scaffolds weighting 415 mg (± 12) containing 4 mg (±0.1) of zoledronic acid (1% in weight) were soaked in 2mL of the PBS solution at 37ºC. Continuous stirring was maintained during the delivery tests and the media was refreshed every time point. After 30 days soaked in PBS, the high deterioration of the scaffolds led us to stop the test. Zoledronic acid released was quantified in the supernatant using a Helios Zeta UV–vis spectrophotometer by reading the absorbance at 209 nm.

*2.5. Culture of human Saos-2 osteoblasts*

Human Saos-2 osteoblasts were cultured in the presence of MBG-PCL or MBG-PCL-zol scaffolds for 7 days in Dulbecco´s Modified Eagle´s Medium with 10% fetal bovine serum, L-glutamine and antibiotics. Cells were then harvested with 0.25% trypsin-EDTA, centrifuged and resuspended in fresh medium for the flow cytometric analysis of cell cycle and apoptosis quantification (see supporting information for further details).

*2.6. Osteoclast differentiation and resorption activity evaluation*

RAW-264.7 macrophages were seeded on glass coverslips in Minimum Essential Medium Eagle Alpha Modification with 5% fetal bovine serum, L-glutamine, antibiotics, 40 ng/mL of mouse recombinant receptor activator for nuclear factor κ B ligand and 25 ng/mL recombinant human macrophage-colony stimulating factor for 7 days. MBG-PCL and MBG-PCL-zol scaffolds were immersed in the medium of osteoclast cultures from either the first day or the sixth day of differentiation until the seventh day. F-actin filaments were stained with rhodamine phalloidin and cell nuclei with 4′-6-diamidino-2′-phenylindole. Then, cells were examined by confocal laser scanning microscopy (see supporting information for further details). The resorption activity of osteoclasts was evaluated on the surface of nanocrystalline hydroxyapatite (nano-HA) disks. Nano-HA disks were prepared by controlled precipitation of calcium and phosphate salts and subsequently heated at temperatures below the sintering point, as previously described [56]. After 7 days of differentiation, cells were detached, and the geometry of resorption cavities produced by osteoclasts on the surface of nano-HA disks was observed by scanning electron microscopy (see supporting information for further details).

*2.7. In vivo studies in osteoporotic sheep model*

This study was approved by our Institutional Ethical Committee following the guidelines of the current normative (Directive 2010/63/EU of the European Parliament and of the Council of September 22, 2010, on the protection of animals used for scientific purposes).

*2.7.1. Induction of osteoporotic model*

Six 4-year-old female Merino sheep (mean preoperative weight of 43.78 ± 5.9 Kg) were included in the study, and all of them were operated on to place as previously reported [57]. Briefly, the sheep were ovarictomiced by lateral laparoscopy, fed with a calcium depleted diet and treated with corticosteroids (see supporting information for further information) .

*2.7.2. Implantation surgical procedure*

Six months after the ovariectomy, the scaffolds were blindly implanted in the sheep under aseptic conditions and the same anaesthetic protocol described elsewhere [57] (see supporting information). The scaffolds were sterilized with formaldehyde vapors, in a steam sterilizer Matachana S.A. for hospital material sterilization and implanted for 12 weeks before being extracted for histological processing. Cylindrical defects (10 x 13 mm in size) were drilled at different locations and one empty defect were left as control (see supporting information for further surgical details)

*2.7.3. Histological processing*

The bone segments containing the defect were dissected out and fixed by immersion in 96% ethanol. Bone segments were divided into 2 segments of 5 mm-thick slices with an automatic stainless-steel special saw for bone, and post-fixed in the same fixative for another 15 days. Half of the bone slices obtained were then fixed in formaldehyde for 3 days and then submerged in a decalcifier solution for 24 hours. (Leica Surgipath decalcifier II). Once decalcified, bone slices were processed and embedded in paraffin in an automated tissue processor (Peloris II Tissue Processor. Leika Biosistems) we made a tissue block with each bone slice. Of each paraffin block we obtained two 5 µm-thick histological sections that were stained with Hematoxylin-Eosin (Agilent Dako Coverstainer for H&E).

*2.7.4. Image acquisition and analysis*

Images were obtained with a Leica DMD1008 and an Olympus BX40 microscopes and analysed with ImageJ 1.x to calculate defect area and bone ingrowth area in each section, as well as number of blood vessels. Histomorphometric quantification of bone ingrowth, inflammatory component and blood vessels density (table 1) was carried out by a blind reviewer (MSP). For histological evaluation, three sections of each sample were evaluated.

For the evaluation of inflammatory component, a five graded scale based on the density and distribution (absent/mild/moderate/marked/ very marked, scored 0 to 4) was used. In the case of extent of vascularization, we used a four graded scale based on the density of blood vessels (absent/mild/moderate/marked/, scored 0 to 3) of vascularity. Initially, we scanned the H&E slide at low power magnification (x4), we identified the 3 hot spots with the higher vascular density, and we counted the number of vessels per 20X magnification field. (Olympus BX40).

*2.8. Statistics*

Data are expressed as means + standard deviations of a representative of three repetitive experiments carried out in triplicate. Statistical analysis was performed by using the Statistical Package for the Social Sciences (SPSS) version 22 software. Statistical comparisons were made by analysis of variance (ANOVA). Scheffé test was used for *post hoc* evaluations of differences among groups. In all statistical evaluations, $p < 0.05$ was considered as statistically significant.

# 3. Results

*3.1. Physicochemical characterization*

SEM micrographs (Figures 1a and b) show cylindrical scaffolds with a regular porous tetragonal structure with pores of 1 mm in size and struts of 400-500 μm in thickness. Thermogravimetric analysis indicated 39 (±2) % in weight of ε-polycaprolactone and FTIR spectra showed the corresponding absorption bands of both MBG and polymer (Figure S1 in supporting information)

*3.2 In vitro drug release test.*

The *in vitro* release of zoledronic acid from MBG-PCL-zol scaffolds was followed for 30 days (Figure 2). The release profile shows a slow delivery during the first 27 days, reaching 28 % of drug content. After this period, the zoledronic acid release is fostered, reaching values above 50 % after 30 days in PBS.

The changes occurred on the scaffolds surface during the test were followed by SEM (Figure 3). Before soaking in PBS (Figure 3a), the surface of the struts appears coated by a continuous phase assignable to ε-PCL, which covers a second discrete phase (MBG particles). Only occasionally, MBG particles are exposed through the very scarce porosity of the struts surface. After 21 days (Figure 3b), the surface appearance has significantly changed. The micrograph shows more porosity and evidences that most of the continuous phase at the surface has disappeared, thus allowing the MBGs particles to be in contact with the surrounding fluid. After 30 days (figure 3c), the extent in which polymer has degraded is higher and leads to detachments at the contact points between rods (figure 3c), where the broken fibres of ε-PCL can be observed.

The $N_2$ adsorption-desorption isotherms profile (Figure 3d) and textural parameters (table 2) indicate that MBG-PCL-zol scaffolds has a very low surface area and porosity. These data agree with the observations made by SEM, suggesting that most of the porosity of MBG-58S is coated by ε-PCL. After 30 days in PBS, $N_2$ isotherms profiles changed associated with to an increase of surface and porosity.

*3.3. In vitro cell culture tests*

Figure 4 shows the percentages of cells within each cycle phase in all these conditions evidencing that no alterations were detected in the $G_0/G_1$ and S phases in the presence of MBG-PCL or MBG-PCL-zol scaffolds. However, MBG-PCL-zol scaffolds induced a significant increase ($p < 0.005$) of $SubG_1$ fraction (apoptotic cells) and a significant decrease of $G_2/M$ phase ($p<0.005$). These results evidence that zoledronic acid released from MBG-PCL-zol samples produced osteoblast apoptosis accompanied by a delay of osteoblast proliferation, which is reflected by the decrease of $G_2/M$ percentage.

Figure 5 shows the effects of MBG-PCL and MBG-PCL-zol scaffolds on the morphology of osteoclast-like cells observed by confocal microscopy after 7 days of differentiation. Multinucleated cells and actin rings were observed in the absence of scaffolds (Control in Figure 5a) and when MBG-PCL scaffolds were immersed in the culture medium from either the first day (Figure 5b) or the sixth day (Figure 5c) of differentiation until the seventh day, revealing osteoclast-like cell differentiation from RAW macrophages after 7 days in these conditions. Concerning MBG-PCL-zol scaffolds, a time dependent effect was observed. The presence of MBG-PCL-zol scaffolds from the first day of differentiation (Figure 5d) until the seventh day induced cell damage and osteoclast-like cells were not found in these cultures. However, when MBG-PCL-zol scaffolds were immersed in the medium from the sixth day of differentiation (Figure 5e) until the seventh day, multinucleated cells with actin ring were observed, thus evidencing that the presence of these scaffolds after the sixth day did not inhibit the osteoclastogenesis.

The resorption cavities left by osteoclast-like cells on nanocrystalline hydroxyapatite disks were observed by SEM (Figure S2 in supporting information). Cavities of 20 to 30 micrometers in size were observed in the absence of scaffolds (Control) and in the presence of both MBG-PCL and MBG-PCL-zol scaffolds. The size of these cavities agrees with the size of the actin rings previously observed (Figure 5).

*3.4. Histological evaluation and histomorphometric quantification of bone ingrowth*

Histological examinations of bones after MBG-PCL implantation show the presence reparative trabecular bone from the peripheral host bone but also within the macroporous structure of the scaffold (Figure 6a). At higher magnification (Figure 6b) the histological observation of MBG-PCL also shows the presence of osteoblastic borders surrounding the newly-mineralized tissues, endothelial cells forming new vessels, multinucleated cells identified as osteoclasts and bone marrow. In the case of MBG-PCL-zol, the histological observation showed the infiltration of fibrotic tissue (figure 6c) and a large amount of inflammatory component associated to the scaffold surface (figure 6d).

Figure 7a shows the extent in which new bone formation occurs in the control defect as well as in the presence of both type of scaffolds. In control defect, no significant new bone formation could be observed in the inner part of the defect (Figure S5 in supporting information), pointing out that bone regeneration is not possible after 12 weeks in the osteoporotic model used in this work. The implantation of MBG-PCL scaffolds not only provokes a statistically significant increase in bone formation at the defect site, but also a significant increase of the trabecular thickness (figure 7b). In the case MBG-PCL-zol scaffolds, the ossification of the defect is almost equivalent to that observed for control defects in both bone volume formation and trabecular thickness, i. e. only new bone formation appeared in the gap between the host bone and the implant, whereas the inner part of the scaffold is colonized by fibrous tissue.

The scaffolds were scored for the presence of osteoblasts and osteoclasts within their pores, as well as for inflammatory response and vascularization (table S1 in supporting information). Figure 8a shows the scoring results for the presence of osteoblasts within the scaffold pores. MBG-PCL showed very high score (> 3.5), which was significantly higher than MBG-PCL-zol (around 2). The scoring of osteoclast presence within the scaffold pores is shown in Figure 8b. At 12 weeks there was no difference between MBG-PCL and MBG-PCL-zol despite of the presence of zoledronic acid in the latter. Scoring of inflammatory response (Figure 8c) showed MBG-PCL-zol scaffolds to elicit an intense response (> 3), significantly higher compared with the low inflammatory response observed for MBG-PCL (around 1). Scoring of vascularization showed no significant differences in the scaffolds (Figure 8d).

## 4. Discussion

In this work we have designed and manufactured macroporous scaffolds by robocasting. SEM observations indicate that robocasting allows for obtaining scaffolds made of MBG and ε-PCL in an accurate and reproducible manner. In this composite system, MBG provides the osteogenic properties to the scaffolds, whereas PCL acts as a binder during the scaffolds manufacture and reduces the brittleness of the pure MBG scaffolds. For this reason, our aim was to introduce the maximum amount of MBG but keeping the rheological characteristics of the injecting paste and preserving mechanical properties strong enough to handle the scaffolds during the surgery. This ratio was MBG:PCL 60:40 used in this work.

SEM observations show that MBG-PCL scaffolds have a surface of continuous and smooth appearance. This fact could be due to the presence of a higher amount of polymeric ε-PCL, that covers MBG particles and occludes the intrinsic porosity of this bioactive glass. This heterogeneity in components distribution has been previously observed by Yun et al [35] and could be explained in terms of the kinetic evaporation of dichlorometane. Since ε-PCL solidifies insofar the solvent is evaporated, ε-PCL placed at the liquid-air interface solidifies faster than PCL entrapped within the inner part of the struts, leading to a phase separation with a PCL rich layer at the surface, as it can be seen in our SEM observation and confirmed by the dramatic surface reduction and porosity occlusion determined by $N_2$ adsorption analysis.

ε-PCL is a relatively stable polymer in aqueous media due to its high degree of crystallinity and hydrophobicity. Its higher presence at the surface delays the drug release so that zoledronic acid is released very slowly during the first four weeks to the surrounded media. Once this external layer is partially degraded, zoledronic acid release is fostered.

The release of zoledronic acid has a considerable influence on osteoblast-like cells and osteoclasts. Our cell culture studies indicate that the incorporation of zoledronic acid in these scaffolds induced a significant increase of apoptosis in osteoblast-like cells and inhibits osteoclastogenesis, although this last effect could be only observed when osteoclast precursor cells were in contact with MBG-PCL-zol scaffolds from the first day of study. This fact can be explained in terms of the release profile of zoledronic acid from MBG-PCL-zol scaffolds. The drug release test in PBS showed that, after 24 hours in PBS, less than 2% of zoledronic acid had been released. In contact with osteoclast precursor cells (macrophages), this amount of drug seems to be not enough for inhibiting the differentiation towards osteoclast cells. After six days in contact with MBG-PCL-zol the amount of drug released (about 12 % of the payload) seems to be enough for damaging macrophages and for avoiding their differentiation to osteoclasts. However, although the presence of MBG-PCL-zol scaffolds from the first day until the seventh day induced cell damage on macrophages, the resorptive activity observed in this condition suggests that those macrophages that could survive were able to differentiate into osteoclast-like cells.

The *in vivo* studies reveal that MBG-PCL scaffolds promotes bone regeneration in osteoporotic sheep, reaching higher ossification volume after 12 weeks and much larger trabecular thickness compared to the non-treated control defect. MBG-PCL scaffolds stimulated bone remodeling by providing a porous architecture that allows the formation of new bone, through the recruitment of osteoblasts even in the inner sites of the scaffolds. In addition, after the degradation of the external ε-PCL, the MBG material is exposed to the defect environment and could exert the osteogenic properties addressed by several authors [8,18] even under osteoporotic conditions. However, the regenerative properties of MBG-PCL scaffolds are lost after the incorporation of zoledronic acid.

Our experiments demonstrate that zoledronic acid cannot be used as adjuvant therapy to bone grafts for the improvement of bone regeneration in skeletal defects. Contrarily to the beneficial effects described by other authors [42,44,45], in our case the local release of zoledronic acid elicits an intense inflammatory and fibrotic response, which impedes the formation of a calcified afibrilar phase required for new bone formation on implant surfaces. Certainly, Peter et al. [42] reported that the local release from the surface of non-porous titanium implants improves the osseointegration, as zoledronic acid increases the BDM of the surrounding host bone. In this case, zoledronic acid partially inhibits the osteoclast activity allowing this increase of BMD. However, the incorporation of zoledronic acid is deleterious for the regenerative properties of our MBG-PCL macroporous scaffolds. Our *in vitro* studies evidenced that it also affects to osteoblast like cells and would be also in agreement with the decrease of bone formation markers previously reported by Delmas et al [49]. In our experimental conditions, negative effects have been observed in both osteoclasts and osteoblasts due to zoledronic acid release from MBG-PCL-zol scaffolds. Recent studies show that this agent at micromolar doses produces negative effects on osteoblast viability, mineralization and collagen synthesis [58]. Concerning osteoclasts, it is known that zoledronic acid inhibits osteoclast proliferation and induces osteoclast apoptosis through the farnesyl diphosphate-mediated mevalonate pathway and caspase activation [59]. However, osteoclast precursors may develop resistance to zoledronic acid induced apoptosis through p38 MAPK-mediated pathway. It has been observed an enhanced effect of zoledronic acid on increasing the bone mineral density of ovariectomized mice as a consequence of inhibiting this pathway [59]. Concerning the doses of zoledronic acid used by other authors *in vitro,* osteoblasts were treated with 0.01-10 µM zoledronic acid [58] and osteoclasts were treated with 50-200 µM zoledronic acid [59]. In our studies, each MBG-PCL-zol scaffold contains about 4 mg of zoledronic acid, and the maximum concentration reached would be 2 g/L in the case of total drug release, which corresponds to a concentration of 700 µM zoledronic acid (MW 272.09). However, it is important to take into account that in our *in vitro* studies, after 7 days of culture, the zoledronic acid release from the scaffold into the culture medium is about 13% (Figure 2), which corresponds to 91 µM zoledronic acid. This amount is in the range of the doses used by other authors for *in vitro* studies with bone cells [58,59]. The drastic effects produced by MBG-PCL-zol scaffolds on osteoclasts could be due to an inhibition of p38 MAPK-mediated pathway and/or to the activation of the farnesyl diphosphate-mediated mevalonate pathway and caspase activation, although further experiments would be required to confirm this point.

Our *in vivo* results indicate that local concentration of zoledronic acid is very high in the defect site. This could be due to an excessive amount of drug incorporated to the scaffolds, but also due to the macroporous architecture of the scaffolds that favors the local drug accumulation at the defect site. Regarding the amount of drug, each MBG-PCL-zol scaffold contains about 4 mg of zoledronic acid, which is a common dosage for intravenously infusions in humans. In this sense the loading of zoledronic acid in our scaffolds is not expected to have such a dramatical effects on sheep weighting around 50 Kg. Moreover, Verron et al [45] also observed bone augmentation and trabeculae reinforcement using a very similar animal model (osteoporotic sheep) and a much higher zoledronate amount (9.2 mg per implant) incorporated to an injectable mixture of calcium phosphate and cellulosic-derived hydrogel. Therefore, doses of 4 mg should not be very high according to other studies and the reasons for this local overdose could be found in the macroporous architecture of the scaffolds. In the case of non-porous implants, drug diffuses from the implant surface towards the surrounding host tissue. Insofar drug is released, a concentration gradient occurs between the implant surface and the peripheral tissues, in such a way that drug concentration decreases with the distance to the implant. Interestingly, it has been reported that bone tissue in direct contact with implants containing high dosage of zoledronic acid exhibited lower quality, whereas trabeculae at 200 μm away from the implant showed larger trabeculae thickness [44]. In the case of our macroporous scaffolds implanted in a hollow defect, zoledronic acid is not only released towards the surrounding tissues but also inwards the hollow defect, so that local concentration increases insofar the drug is released. The local overdose reached in this situation would explain the intense inflammatory response observed in our experiments for MBG-PCL-zol scaffolds. This inflammatory response agrees with the arthralgias observed in some patients as side effects or because of overdosage with zoledronic acid [60]. Summarizing, whereas implantation of MBG-PCL macroporous scaffolds is an excellent strategy for bone regeneration even in osteoporotic environment, the incorporation of zoledronic acid is a counter-productive strategy. This is due to the partial inhibition of bone formation associated to zoledronic acid together with the porous architecture of the MBG-PCL scaffolds, which facilitates the local overdose of the drug within the defect. While this strategy can be useful to improve the fixation of non-porous orthopedic devices, it must be discarded for bone grafts designed for regeneration purposes.

## 5. Conclusions

MBG-PCL macroporous scaffolds have been prepared by robocasting for bone regeneration purposes. These scaffolds exhibit excellent biocompatibility under both *in vitro* and *in vivo* conditions. In this sense, MBG-PCL scaffolds stimulate bone regeneration in a scenario of osteoporosis. These implants promote new bone formation even when the site morphological and metabolic conditions are not favorable to osteointegration and subsequent regeneration.

In order to evaluate the effects of the local release of an antiresorptive drug in bone defects, zoledronic acid was added to the scaffolds. The drug inhibits macrophage differentiation into osteoclasts but also induced apoptosis in osteoblast-like cells. Under *in vivo* conditions, the incorporation of zoledronic acid inhibited the new bone formation and promoted a strong inflammatory response.

## Acknowledgements

This study was supported by research grants from the Ministerio de Economía y Competitividad (project MAT2016-75611-R AEI/FEDER, UE). M.V.-R. acknowledges funding from the European Research Council (Advanced Grant VERDI; ERC-2015-AdG Proposal 694160). The authors wish to thank the ICTS Centro Nacional de Microscopia Electrónica (Spain) and Centro de Citometría de Flujo y Microscopía de Fluorescencia of the Universidad Complutense de Madrid (Spain) for their technical assistance.

**Figure captions**

**Figure 1.** Scanning electron micrographs of MBG-PCL-zol scaffolds. (a) magnification x 7 and (b) magnification x 20.

**Figure 2.** Zoledronic acid release from MBG-PCL-zol as a function of soaking time.

**Figure 3.** Scanning electron micrographs of MBG-PCL-zol scaffolds before (a) and after 21 days (b) and 30 days (c) of drug release test. Nitrogen adsorption/desorption isotherms of MBG-PCL-zol scaffolds before and after 30 days soaked in PBS (d).

**Figure 4.** Effects of MBG-PCL (grey) and MBG-PCL-zol (black) scaffolds on cell cycle phases of human Saos-2 osteoblasts after 7 days of culture. Controls in the absence of scaffolds (white) were carried out in parallel. Statistical significance: *** $p < 0.005$.

**Figure 5**. Effects of MBG-PCL and MBG-PCL-zol scaffolds on the morphology of osteoclast-like cells observed by confocal microscopy after 7 days of differentiation. (a) Controls in the absence of scaffolds; (b) MBG-PCL scaffolds immersed in the medium of osteoclast cultures from the first day; (c) MBG-PCL scaffolds immersed in the medium of osteoclast cultures from the sixth day; (d) MBG-PCL-zol scaffolds immersed in the medium of osteoclast cultures from the first day; (e) MBG-PCL-zol scaffolds immersed in the medium of osteoclast cultures from the sixth day. Actin was stained with rhodamine-phalloidin (red) and cell nuclei with DAPI (blue).

**Figure 6.** Histological examination after 12 weeks of implantation of MBG-PCL (a and b) and MBG-PCL-zol (c and d). The inset in figure 6.b shows osteoblast border (1), blood vessel (2) and osteoclast cells (3). (*) indicates inflammatory component.

**Figure 7.** Histomorphometrical studies of bones of osteoporotic sheep. (a) Ossification volume, (b) trabeculae thickness. Statistical significance * $p < 0.005$

**Figure 8.** Results of histological scoring. Scores for (a) the presence of osteoblasts at the scaffolds porosity, (b) the presence of osteoclasts at the scaffolds porosity, (c) the amount of inflammatory component and (d) the extent of vascularization degree. Statistical significance * $p < 0.005$.